\documentclass[12pt]{article}
\usepackage{graphics}
\usepackage{cite}
\usepackage{epsfig}

\newcommand{\beq}{\begin{equation}}
\newcommand{\eeq}{\end{equation}}
\newcommand{\beqa}{\begin{eqnarray}}
\newcommand{\eeqa}{\end{eqnarray}}
\newcommand{\bea}{\begin{eqnarray}}
\newcommand{\eea}{\end{eqnarray}}
\newcommand   {\ev}[1]   {\langle #1\rangle}
\newcommand   {\Ca}      {C${}_{\alpha}$}
\newcommand   {\Cb}      {C${}_{\beta}$}
\newcommand   {\Cp}      {C${}^{\prime}$}
\newcommand   {\Eloc}    {E_{\mbox{{\scriptsize loc}}}}
\newcommand   {\Cv}      {C_{\mbox{{\scriptsize v}}}}
\newcommand   {\Esa}     {E_{\mbox{{\scriptsize sa}}}}
\newcommand   {\Ehb}     {E_{\mbox{{\scriptsize hb}}}}
\newcommand   {\Eaa}     {E_{\mbox{{\scriptsize AA}}}}
\newcommand   {\ephi}    {\epsilon_\phi}
\newcommand   {\epsi}    {\epsilon_\psi}
\newcommand   {\esa}     {\epsilon_{\mbox{{\scriptsize sa}}}}
\newcommand   {\ehb}     {\epsilon_{\mbox{{\scriptsize hb}}}}
\newcommand   {\tehb}    {\tilde\epsilon_{\mbox{{\scriptsize hb}}}}
\newcommand   {\eaa}     {\epsilon_{\mbox{{\scriptsize AA}}}}
\newcommand   {\teaa}    {\tilde\epsilon_{\mbox{{\scriptsize AA}}}}
\newcommand   {\shb}     {\sigma_{\mbox{{\scriptsize hb}}}}
\newcommand   {\saa}     {\sigma_{\mbox{{\scriptsize AA}}}}
\newcommand   {\QFU}     {Q_{\mbox{{\scriptsize FU}}}}
\newcommand   {\QBU}     {Q_{\mbox{{\scriptsize BU}}}}
\newcommand   {\dBU}     {\delta_{\mbox{{\scriptsize BU}}}}
\newcommand   {\dFU}     {\delta_{\mbox{{\scriptsize FU}}}}

\newcommand   {\Tf}      {T_{\mbox{{\scriptsize f}}}}
\newcommand  {\APC}      {{\em Adv. Protein Chem.\ }}

\newcommand  {\COSB}     {{\em Curr.\ Opin.\ Struct.\ Biol.\ }}
\newcommand  {\EL}       {{\em Europhys.\ Lett.\ }}

\newcommand  {\JCP}      {{\em J.\ Chem.\ Phys.\ }}
\newcommand  {\JMB}      {{\em J.\ Mol.\ Biol.\ }}

\newcommand  {\JSP}      {{\em J.\ Stat.\ Phys.\ }}

\newcommand  {\NSB}      {{\em Nature\ Struct.\ Biol.\ }}

\newcommand  {\Pro}      {{\em Proteins:\ Struct.\ Funct.\ Genet.\ }}

\newcommand  {\PNAS}     {{\em Proc.\ Natl.\ Acad.\ Sci.\ USA\ }}
\newcommand  {\PRE}      {{\em Phys.\ Rev.\ E\ }}
\newcommand  {\PRL}      {{\em Phys.\ Rev.\ Lett.\ }}

\newcommand  {\Sci}      {{\em Science\ }}

\newcommand  {\TBS}      {{\em Trends Biochem. Sci.\ }}
\begin{document}                                        

\begin{flushright}
LU TP 00-34\\
Revised version\\
June 2, 2001
\end{flushright}
 
\vspace{0.4in}            

\begin{center}

{\LARGE \bf Hydrogen Bonds, Hydrophobicity Forces}\\
 
\vspace{6pt}
 
{\LARGE \bf and the}\\
 
\vspace{12pt}
 
{\LARGE \bf Character of the Collapse Transition}
 
\vspace{.3in}
                                                
\large
Anders Irb\"ack, Fredrik Sjunnesson and
Stefan Wallin\footnote{E-mail: irback,\,fredriks,\,stefan@thep.lu.se}\\
\vspace{0.10in}
Complex Systems Division, Department of Theoretical Physics\\
Lund University,  S\"olvegatan 14A,  S-223 62 Lund, Sweden \\
{\tt http://www.thep.lu.se/tf2/complex/}\\
 
\vspace{0.3in}
 
Contribution to {\it Proceedings of the ISI Workshop
``Protein Folding: Simple Models and Experiments''},
Torino, April 27 - May 2, 2000.
\end{center}
 
\vspace{0.2in}
\normalsize                            
Abstract:\\
We study the thermodynamic behavior of a model protein with 54 amino acids
that is designed to form a three-helix bundle in its native state. The model 
contains three types of amino acids and five to six atoms per amino acid, 
and has the Ramachandran torsion angles as its only degrees of freedom.
The force field is based on hydrogen bonds and effective hydrophobicity 
forces. We study how the character of the collapse transition 
depends on the strengths of these forces. For a suitable choice 
of these two parameters, it is found that the collapse transition is  
first-order-like and coincides with the folding transition.  
Also shown is that the corresponding one- and 
two-helix segments make less stable secondary structure than the   
three-helix sequence.

\newpage

\section{Introduction}  

The study of the formation of the native structures of proteins  
is hampered by computational limitations 
and uncertainties about the relevant forces, 
which makes model building a delicate and highly relevant task. 
Most current models use one or both of two quite drastic 
approximations, the lattice and G\=o~\cite{Go:78} 
approximations, where the latter amounts to ignoring interactions 
that do not favor the desired structure. Models of these types have 
provided valuable insights into the physical principles of protein 
folding~\cite{Sali:94,Bryngelson:95,Dill:97,Klimov:98,Nymeyer:98},
but have their obvious limitations.  

Besides being computationally convenient, lattice models have  
the important advantage that it is known what is needed in 
order for stable and fast-folding chains to exist; it can be 
achieved by using a simple contact potential. For off-lattice models this 
is largely unknown, although one thing that seems clear is that it is not 
enough to simply use a potential analogous to the contact 
potential~\cite{Socci:94,Irback:97,Vendruscolo:99}. 
Because of this uncertainty, and because of evidence that 
the native structure in itself is a major determinant of folding 
kinetics~\cite{Alm:99}, many off-lattice studies have been 
based on G\=o-type potentials.
        
In this paper, we take a different approach, by discussing an 
off-lattice model, proposed in~\cite{Irback:00}, that does not use the G\=o 
approximation. In this model, the formation of 
a native structure is driven by hydrogen bonding and 
effective hydrophobicity forces.  The model has 
three types of amino acids and the Ramachandran angles $\phi_i$ and 
$\psi_i$~\cite{Ramachandran:68} as its degrees of freedom. 
Each amino acid is represented by five or six atoms. 

Using this model, we study a 
three-helix-bundle protein with 54 amino acids, 
which represents a truncated and simplified version of the 
four-helix-bundle protein {\it de novo} designed by Regan
and Degrado~\cite{Regan:88}.  
To study size dependence, we also look 
at the behavior of the corresponding one- and two-helix 
segments. The thermodynamic properties of these different chains 
are explored by using the method of simulated 
tempering~\cite{Lyubartsev:92,Marinari:92,Irback:95}. 

Two key parameters of this model are the respective strengths
$\ehb$ and $\eaa$ of the hydrogen bonds and hydrophobicity forces.
For a suitable choice of these parameters, to be denoted by 
$(\tehb,\teaa)$, the three-helix sequence is found to have 
the following properties~\cite{Irback:00}:   
\begin{itemize} 
\item It does form a stable three-helix bundle (except for a twofold 
topological degeneracy). 
\item It undergoes a first-order-like folding transition,  
from an expanded state to the native three-helix-bundle state. 
\item It forms more stable secondary structure than the one- and two-helix
segments.  
\end{itemize}
Qualitatively similar results have been obtained previously for 
\Ca~\cite{Nymeyer:98,Shea:98,Shea:99,Clementi:00a,Clementi:00b} 
and all-atom~\cite{Shimada:01} off-lattice chains, but, as far as we 
know, only with G\=o-type potentials.

The paper is organized as follows. In Section~\ref{mod}, we give
a brief description of the model. Our results are presented
in Section~\ref{res}. Here, we first summarize the results obtained
in \cite{Irback:00} for $(\ehb,\eaa)=(\tehb,\teaa)$. We then discuss 
how the character of collapse transition depends on the relative strength 
of $\ehb$ and $\eaa$, by studying the behavior for 
$(\ehb,\eaa)=(\tehb-\kappa,\teaa+\kappa)$ for different $\kappa$.   
We end with a brief summary in Section~\ref{disc}.   

\section{The Model}\label{mod}

\begin{figure}
\begin{center}
\rotatebox{270}{\epsfig{figure=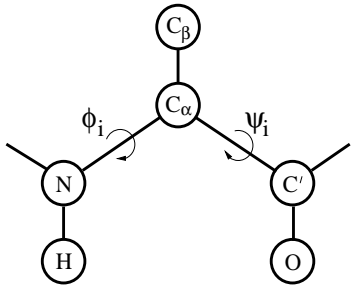,width=3.75cm,height=4.65cm}}
\end{center}
\caption{Schematic figure showing the representation of one amino acid.}  
\label{fig:1}
\end{figure}

The model we study is a reduced off-lattice model.   
Figure~\ref{fig:1} illustrates the representation of one 
amino acid. The side chain is represented 
by a single atom, \Cb, which can be either hydrophobic, 
polar or absent. This gives us three types of 
amino acids: A with hydrophobic \Cb, B with polar \Cb, 
and G (glycine) without \Cb.

The H, O and \Cb\ atoms are all attached to the backbone in a rigid 
way. Furthermore, in the backbone, all bond lengths, bond angles and 
peptide torsion angles ($180^\circ$) are held fixed. 
This leaves us with two degrees of freedom per amino acid, the 
Ramachandran torsion angles $\phi_i$ and $\psi_i$ (see Figure~\ref{fig:1}). 

Our energy function 
\beq
E=\Eloc+\Esa+\Ehb+\Eaa
\label{e}\eeq
is composed of four terms. 
The local potential $\Eloc$ has a standard form with threefold symmetry,
\beq
\Eloc=\frac{\ephi}{2}\sum_i(1 + \cos3\phi_i)
+ \frac{\epsi}{2}\sum_i(1 + \cos3\psi_i)\,.
\eeq
The self-avoidance term $\Esa$ is given by a hard-sphere potential
of the form 
\beq
\Esa=\esa\mathop{{\sum}'}_{i<j}
\bigg(\frac{\sigma_{ij}}{r_{ij}}\bigg)^{12}\,,
\label{sa}\eeq 
where the sum runs over all possible atom pairs except those  
consisting of two hydrophobic \Cb. The hydrogen-bond term $\Ehb$
is given by
\beq
\Ehb= \ehb \sum_{ij}u(r_{ij})v(\alpha_{ij},\beta_{ij})\,,
\label{hb}\eeq
where 
\begin{eqnarray} 
u(r_{ij})&=&  5\bigg(\frac{\shb}{r_{ij}}\bigg)^{12} - 
        6\bigg(\frac{\shb}{r_{ij}}\bigg)^{10}\\
v(\alpha_{ij},\beta_{ij})&=&\left\{ 
        \begin{array}{ll}
 \cos^2\alpha_{ij}\cos^2\beta_{ij} & \ \alpha_{ij},\beta_{ij}>90^{\circ}\\
 0                      & \ \mbox{otherwise}
         \end{array} \right. 
\end{eqnarray}
Here, $i$ and $j$ represent H and O atoms, respectively, 
and $r_{ij}$ denotes the HO distance, $\alpha_{ij}$ the NHO 
angle, and $\beta_{ij}$ the HO\Cp\ angle.  
The last term in Equation~\ref{e} is the hydrophobicity 
energy $\Eaa$, which has the form
\beq
\Eaa=\eaa\sum_{i<j}\bigg[
\bigg(\frac{\saa}{r_{ij}}\bigg)^{12}
-2\bigg(\frac{\saa}{r_{ij}}\bigg)^6\,\bigg]\,,
\eeq
where both $i$ and $j$ represent hydrophobic \Cb. To speed up the simulations, 
a cutoff radius $r_c$ is used,\footnote{The cutoff procedure is
$f(r)\mapsto\tilde f(r)$ where $\tilde f(r)=f(r)-f(r_c)-(r-r_c)f^\prime(r_c)$
if $r<r_c$ and $\tilde f(r)=0$ otherwise.} which is 4.5\AA\ for $\Esa$ 
and $\Ehb$, and 8\AA\ for $\Eaa$.

\begin{figure}
\begin{center}
\epsfig{figure=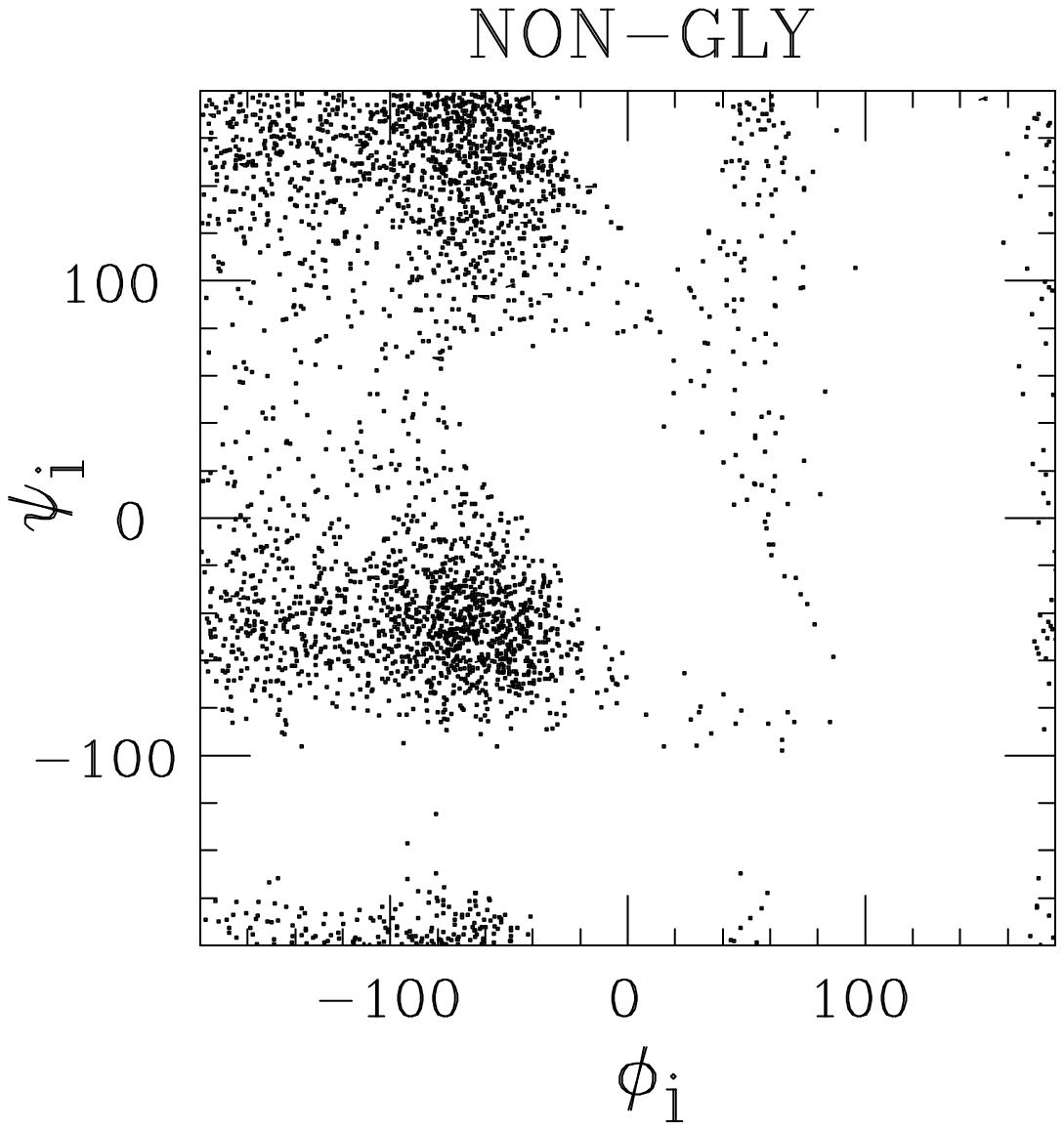,width=3.75cm,height=3.97cm}
\hspace{10mm}
\epsfig{figure=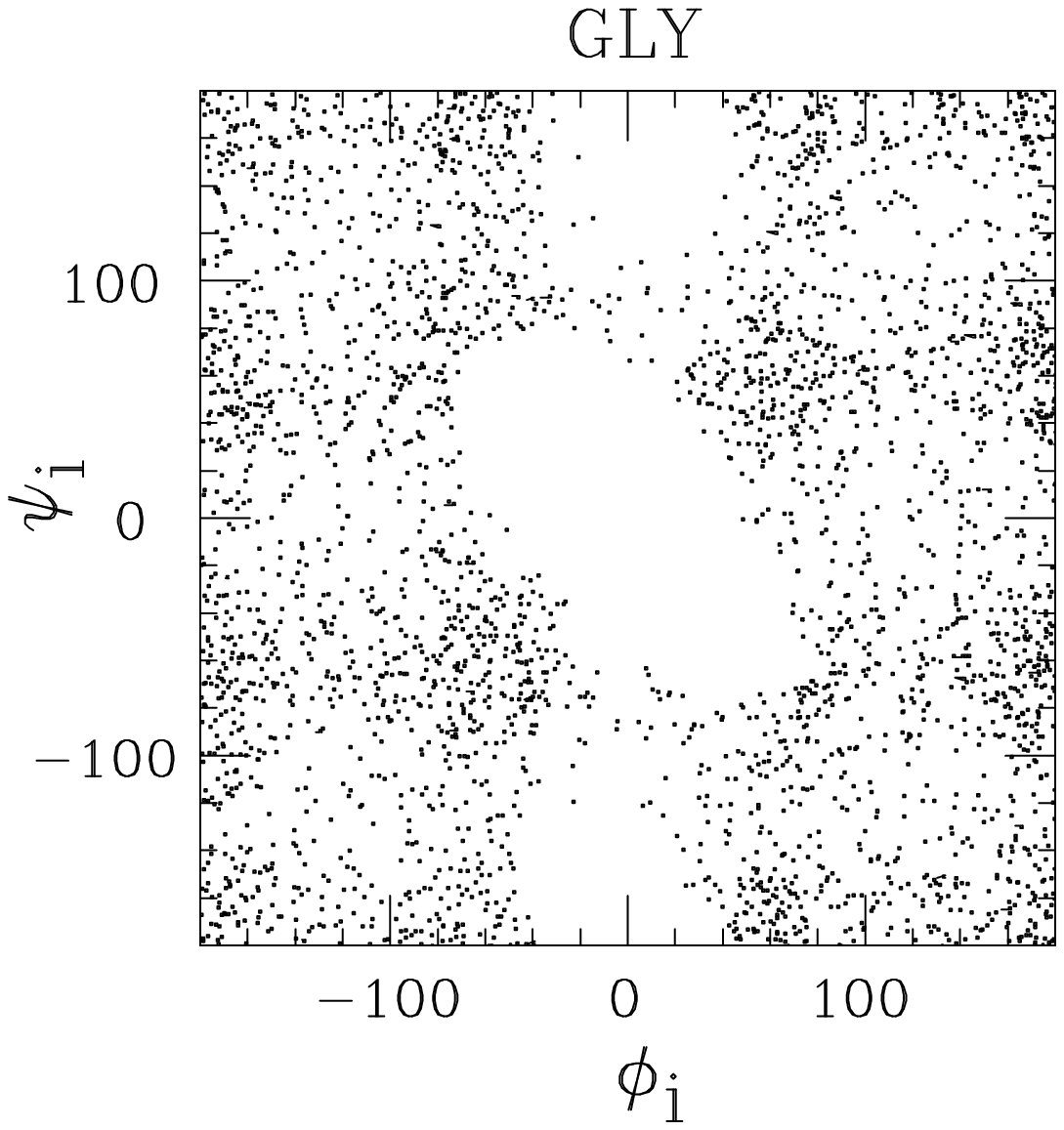,width=3.75cm,height=3.97cm}
\end{center}
\caption{$\phi_i,\psi_i$ scatter plots for non-glycine and glycine, as 
obtained by simulations of the chains GXG for X=A/B and X=G, respectively, 
at $kT=0.625$ (shown is $\phi_i,\psi_i$ for X).} 
\label{fig:2}
\end{figure}

The parameters of the energy function were determined empirically
based on the shape of the Ramachandran $\phi_i,\psi_i$ distribution 
and on the overall thermodynamic behavior of the three-helix-bundle 
protein. A complete list of energy and geometry parameters 
can be found in~\cite{Irback:00}. In Figure~\ref{fig:2}, we show
the final $\phi_i,\psi_i$ distributions for non-glycine (A and B) and 
glycine. 

As mentioned in the introduction, we study the model for different 
$(\ehb,\eaa)$. For $(\ehb,\eaa)=(\tehb,\teaa)=(2.8,2.2)$ 
(dimensionless units), it turns out that
\beq
\tehb/k\Tf\approx4.3\qquad \teaa/k\Tf\approx3.4\,,
\label{para}\eeq   
where $\Tf$ denotes the folding temperature of the three-helix-bundle protein
(see below).

The three sequences studied are listed in Table~\ref{tab:1}. They    
contain 16, 35 and 54 amino acids, respectively. The one-helix 
segment 1H consists of A and B amino acids that are distributed 
in such a way that this segment can form a helix with all hydrophobic 
amino acids on the same side. The three-helix
sequence, 3H, consists of three such stretches of As and Bs 
plus two GGG segments.

\begin{table}
\begin{center}
\begin{tabular}{ll}
\hline
1H:& BBABBAABBABBAABB\\
2H:& 1H--GGG--1H\\
3H:& 1H--GGG--1H--GGG--1H\\
\hline
\end{tabular}
\end{center}
\caption{The sequences studied.} 
\label{tab:1}
\end{table}

There have been several earlier studies of similar-sized helical
proteins using models at comparable levels of 
resolution~\cite{Shea:99,Rey:93,Guo:96,Zhou:97,Koretke:98,Hardin:99,Takada:99}.
Among these studies, most similar to ours is that of
Takada et al.~\cite{Takada:99}. These authors studied the same
sequences, using a somewhat similar chain representation and a 
different, more elaborate force field. 

\section{Results}\label{res}

Using simulated tempering, we study the thermodynamic behavior of the 
chains defined above for 
\beq
\ehb=\tehb-\kappa\qquad\qquad \eaa=\teaa+\kappa
\label{kappa}\eeq
for different $\kappa$. 

\subsection{Balance between Hydrogen Bonds and Hydrophobicity Forces}

We begin with a summary of the results obtained in \cite{Irback:00}
for $\kappa=0$. 

\begin{figure}
\begin{center}
\epsfig{figure=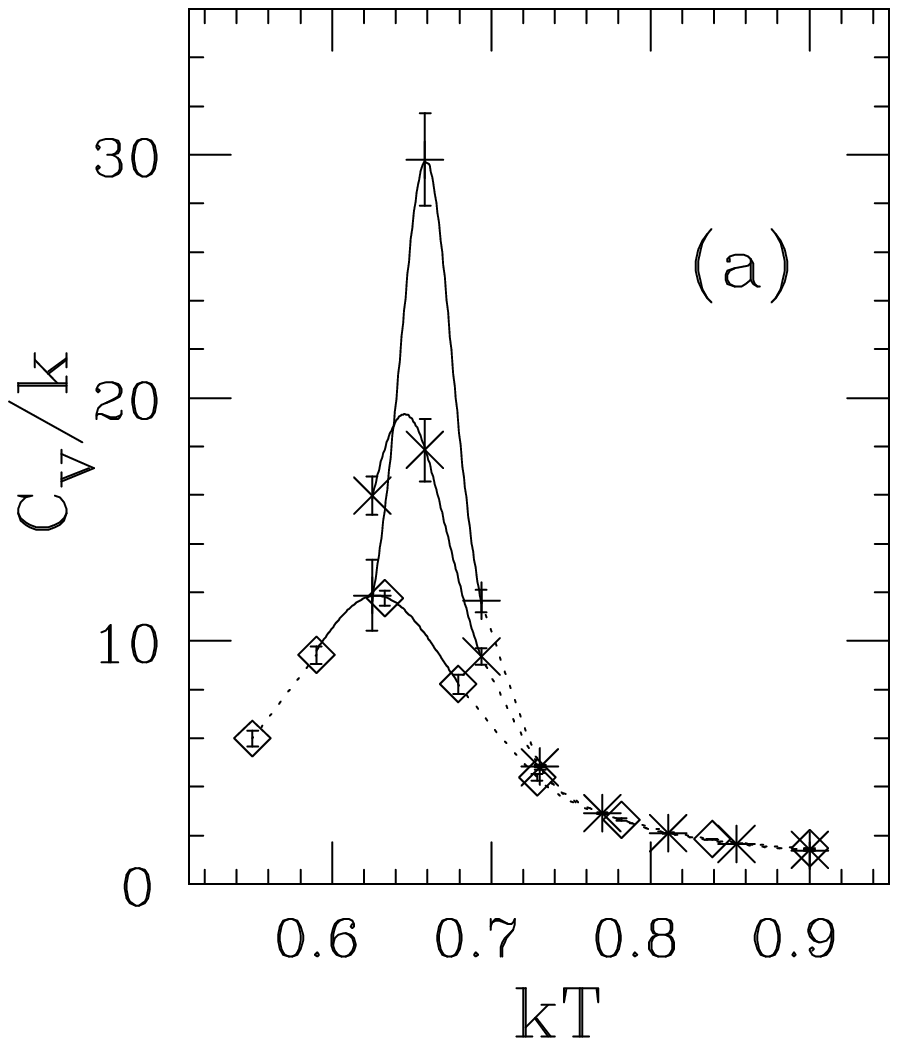,width=3.75cm,height=4.4cm}
\hspace{10mm}
\epsfig{figure=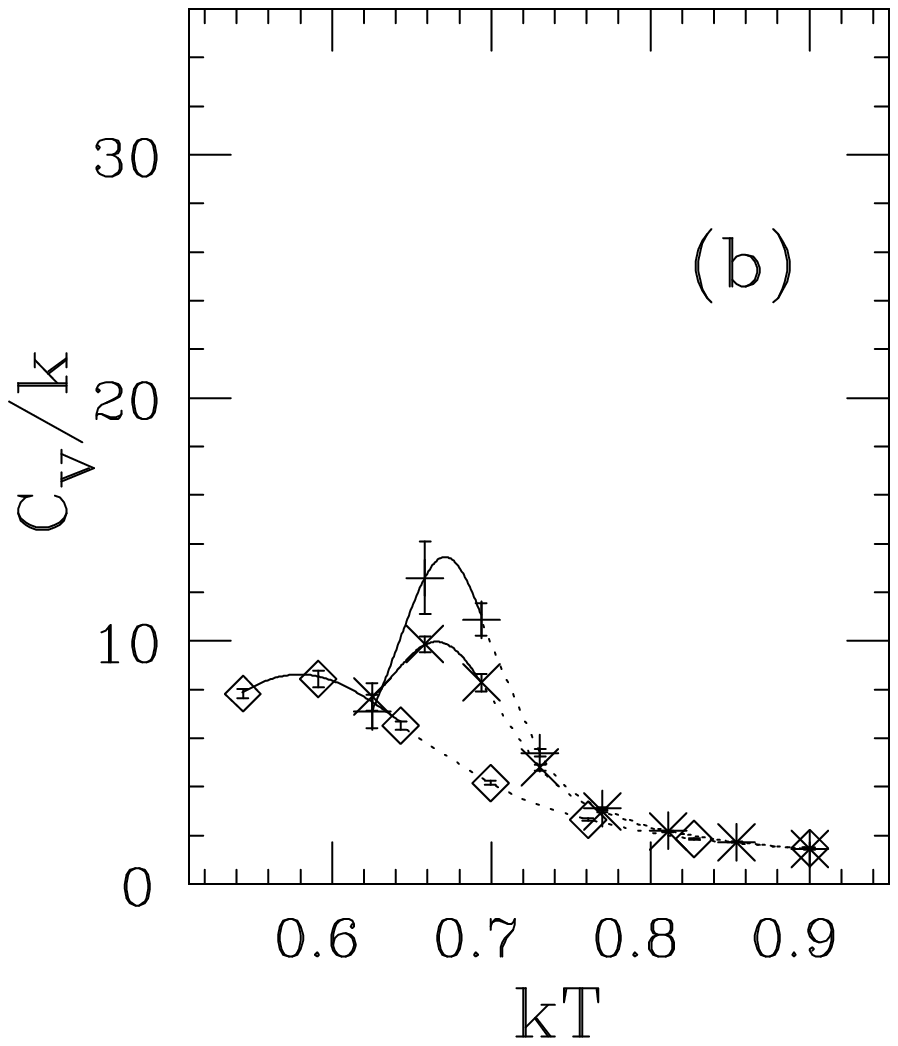,width=3.75cm,height=4.4cm}
\end{center}
\caption{The specific heat $\Cv=(\ev{E^2}-\ev{E}^2)/NkT^2$
against temperature for the sequences 1H ($\diamond$), 
2H ($\times$) and 3H ($+$) (see Table~\protect\ref{tab:1}), 
for (a) $\kappa=0$ and (b) $\kappa=0.3$ ($N$ denotes the
number of amino acids). The full lines represent 
single-histogram extrapolations~\protect\cite{Ferrenberg:88}. Dotted lines are drawn to guide the eye.}
\label{fig:3}
\end{figure}

\begin{figure}
\begin{center}
\epsfig{figure=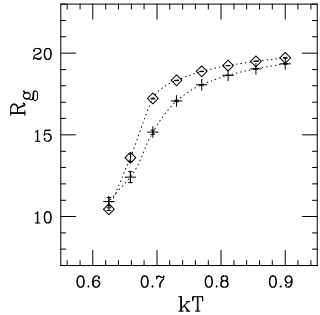,width=4.1cm,height=4.1cm}
\end{center}
\caption{Radius of gyration (in \AA) against temperature for the 
the three-helix sequence, for $\kappa=0$ ($\diamond$) and 
$\kappa=0.3$ ($+$).}
\label{fig:4}
\end{figure}

For this choice of $(\ehb,\eaa)$, it turns out that the three-helix 
sequence exhibits an abrupt collapse transition, 
signaled by a sharp peak in the specific heat. 
This can be seen from Figures~\ref{fig:3}a and \ref{fig:4}, which show the 
specific heat and radius of gyration, respectively, against
temperature. 

\begin{figure}
\begin{center}
\epsfig{figure=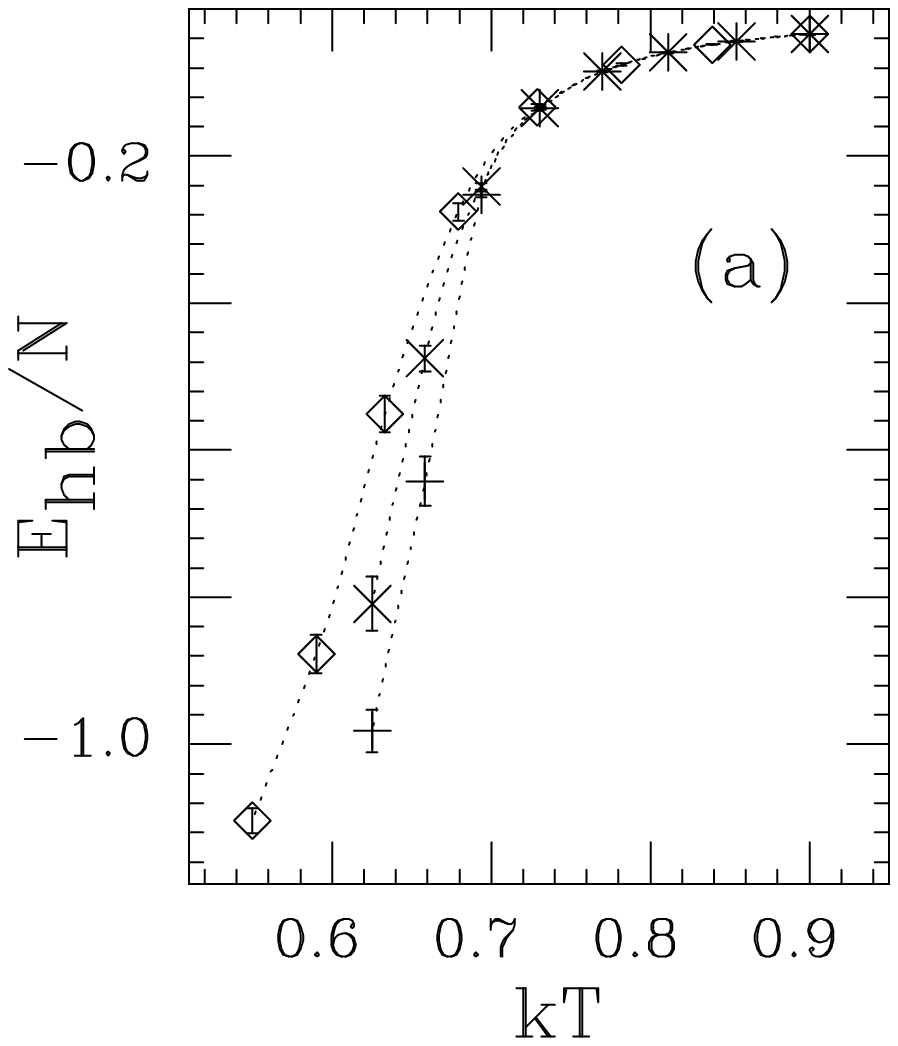,width=3.75cm,height=4.4cm}
\hspace{10mm}
\epsfig{figure=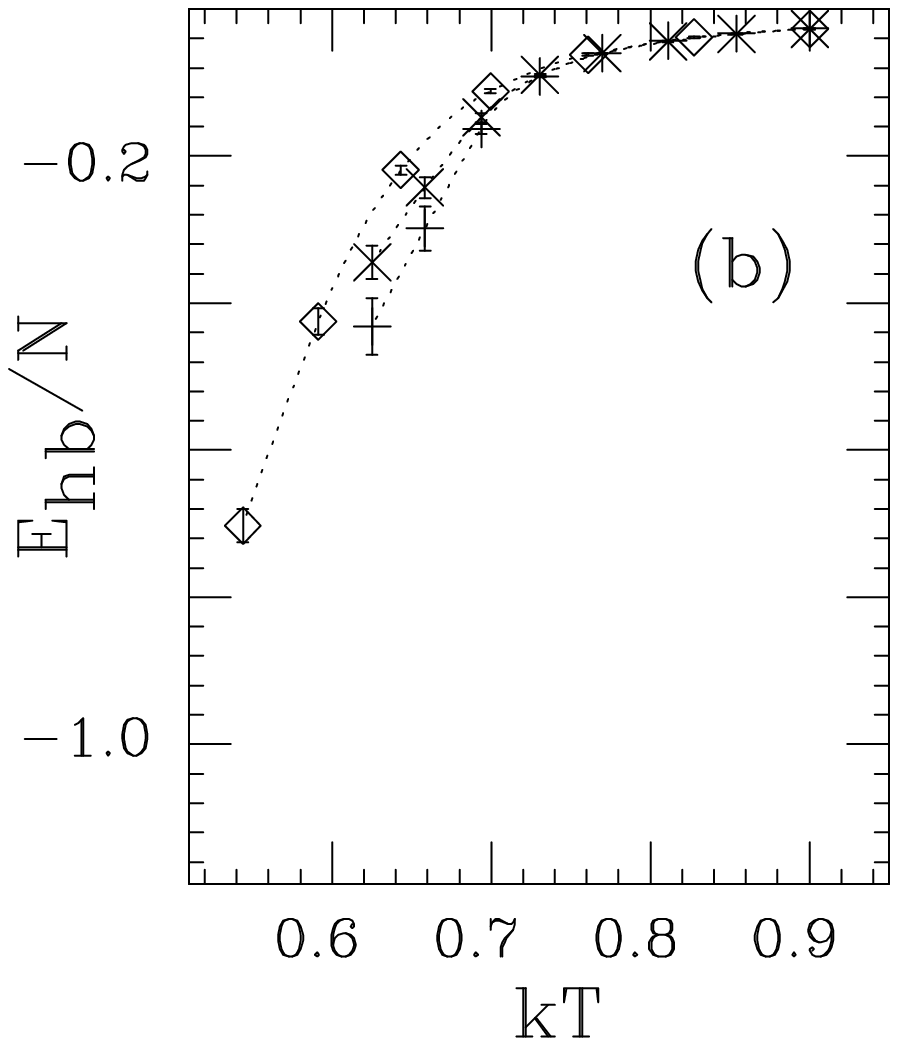,width=3.75cm,height=4.4cm}
\end{center}
\caption{Hydrogen-bond energy per amino acid, $\Ehb/N$,
against temperature for the sequences 1H ($\diamond$), 
2H ($\times$) and 3H ($+$) (see Table~\protect\ref{tab:1}), 
for (a) $\kappa=0$ and (b) $\kappa=0.3$.}
\label{fig:5}
\end{figure}

It is instructive to look at how the results depend on chain length 
near the transition. Two important observations are the following:  
\begin{itemize}
\item The peak in the specific heat gets stronger with increasing 
chain length. The increase in height is not inconsistent with
a linear size dependence, which is what one would expect at a 
conventional first-order phase transition with a latent heat. 
\item The decrease in hydrogen-bond energy per amino acid, $\Ehb/N$, 
with decreasing temperature gets more rapid with increasing chain
length, as shown in Figure~\ref{fig:5}a. This implies that the 
three-helix protein makes more stable secondary structure than 
the one- and two-helix segments.    
\end{itemize}

\begin{figure}
\begin{center}
\epsfig{figure=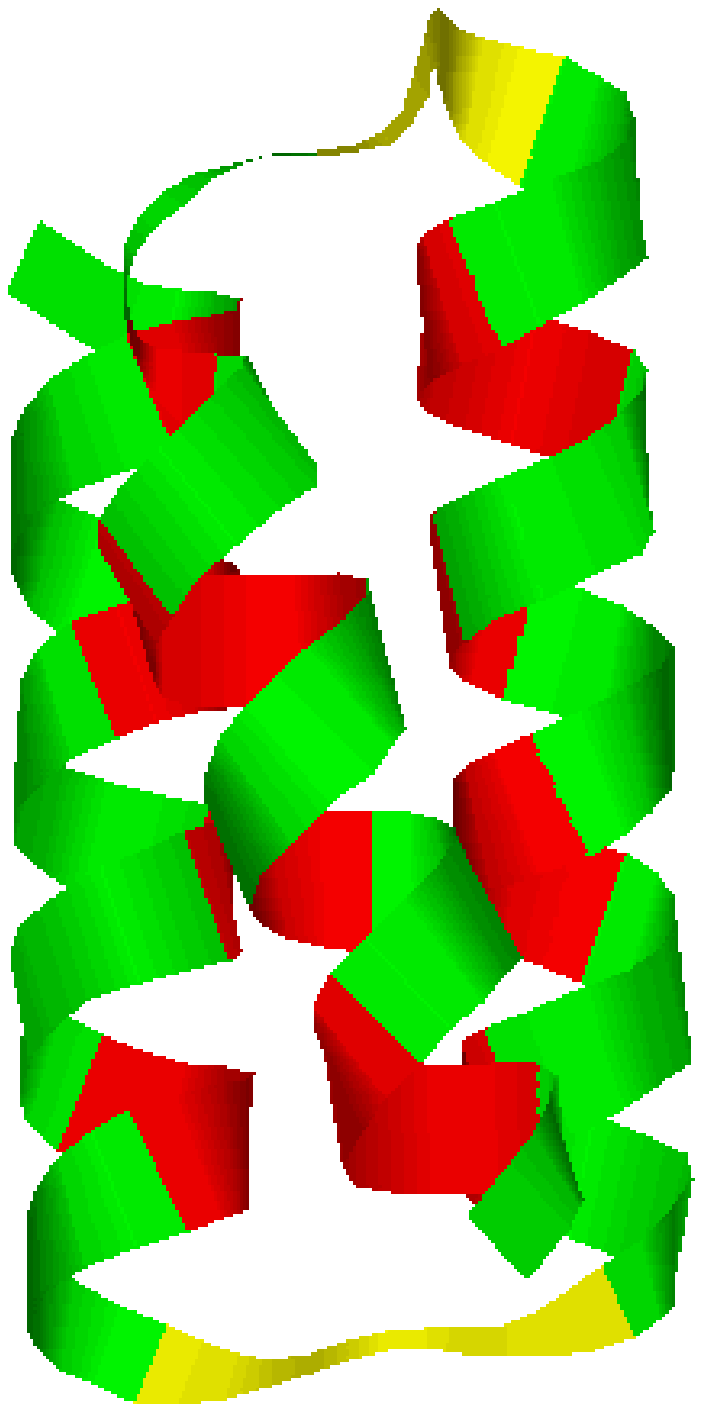,width=1.4cm,height=2.7cm}
\hspace{25mm}
\epsfig{figure=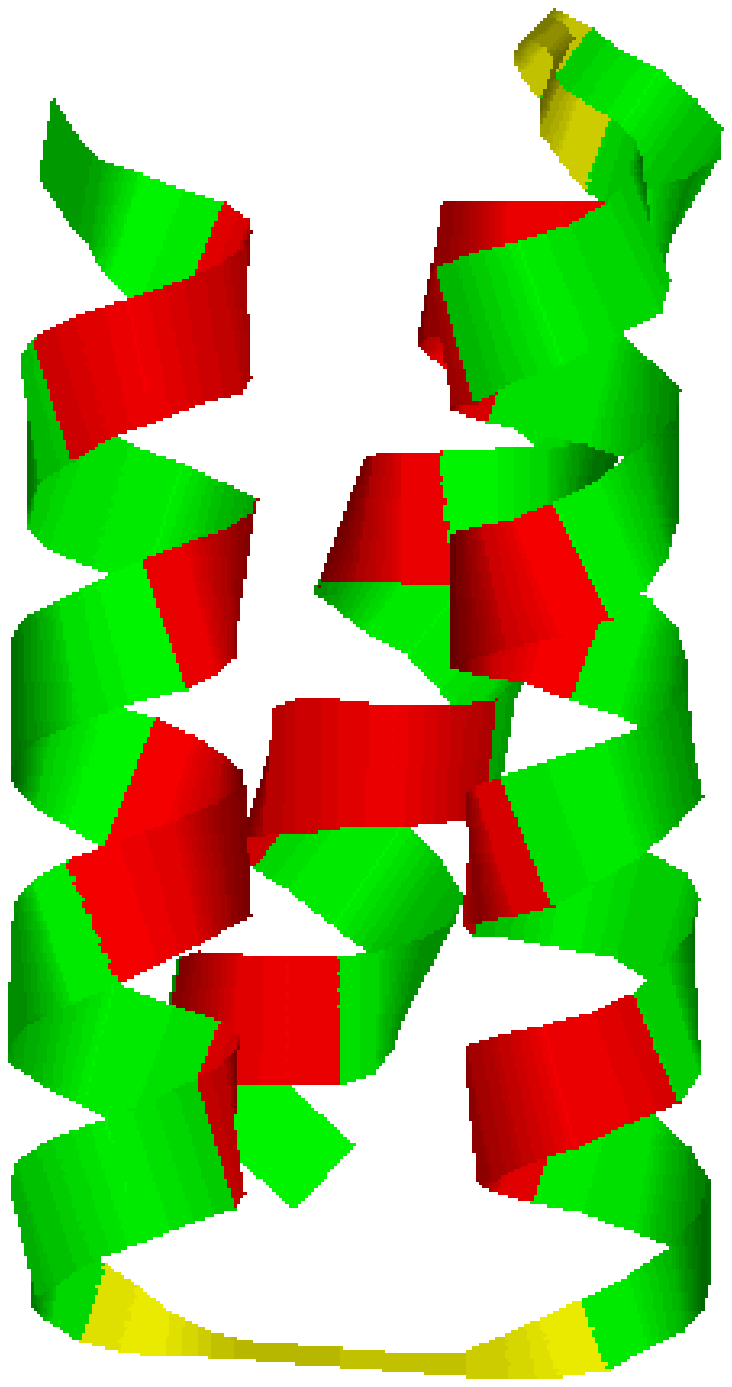,width=1.4cm,height=2.7cm}
\end{center}
\caption{Representative low-temperature structures, FU and BU, respectively.
Drawn with RasMol~\protect\cite{Sayle:95}.}  
\label{fig:6}
\end{figure}

It turns out that the sequence 3H does form a three-helix bundle
at low temperatures. This bundle can have two distinct topologies; if we 
let the first two helices form a U, then the third helix can be either 
in front of or behind that U.  The model is, not unexpectedly, 
unable to discriminate between these two possibilities. To characterize 
low-temperature conformations, we therefore determined two representative 
structures, one for each topology, which, following~\cite{Takada:99}, 
are referred to as FU and BU, respectively. These structures
are shown in Figure~\ref{fig:6}. Given an arbitrary 
conformation, we then measure the root-mean-square 
deviations $\delta_i$ ($i=$\,FU,\,BU) from these two structures (calculated
over all backbone atoms). These deviations are 
converted into similarity parameters $Q_i$ by using  
\beq
Q_i=\exp(-\delta_i^2/100\mbox{\AA}^2)\,.
\label{sim}\eeq

\begin{figure}[t]
\begin{center}
\epsfig{figure=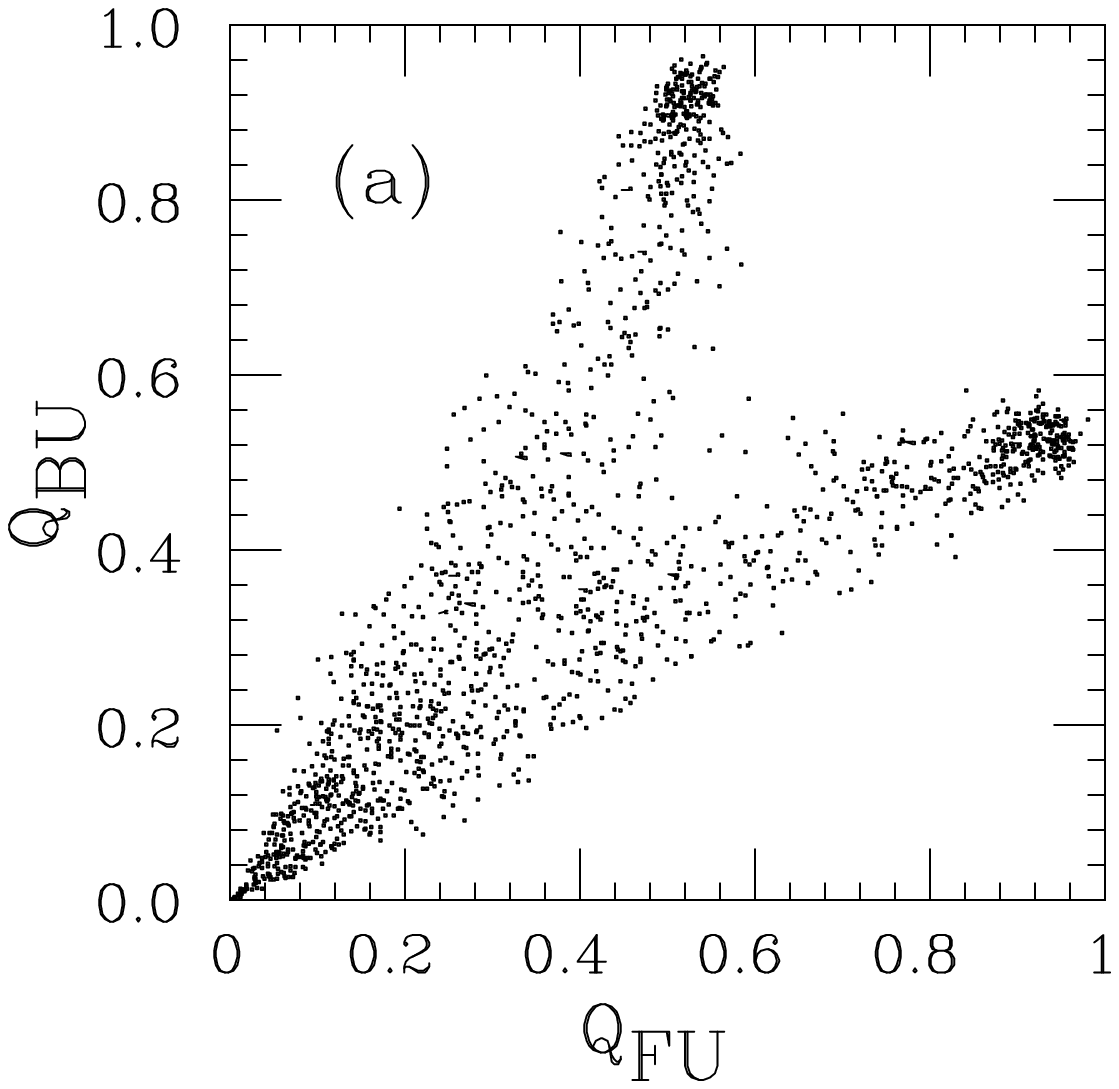,width=3.6cm,height=3.7cm}
\hspace{10mm}
\epsfig{figure=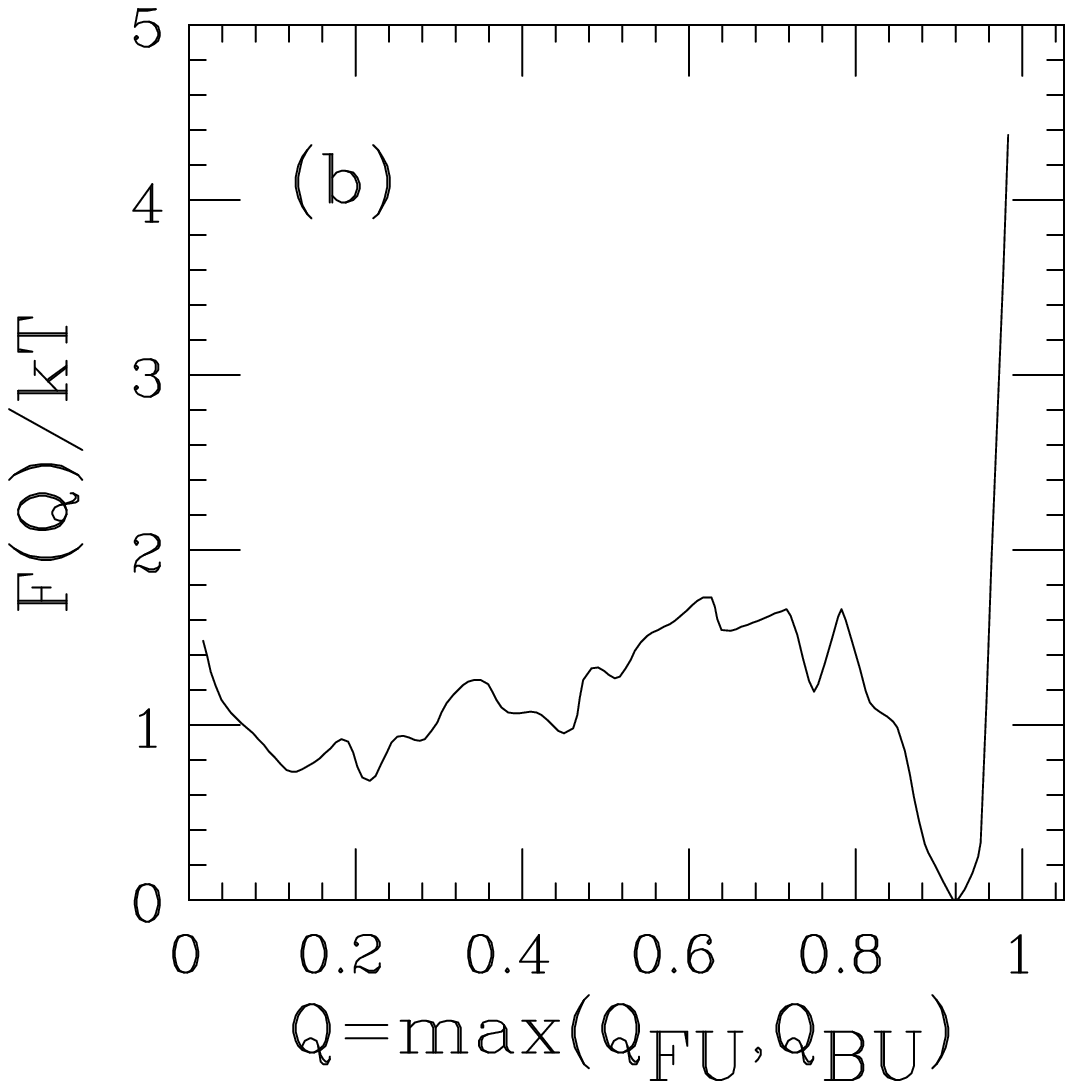,width=3.6cm,height=3.7cm}
\end{center}
\caption{(a) $\QFU,\QBU$ scatter plot (see Equation~\protect\ref{sim})  
at the collapse temperature. 
(b) Free-energy profile $F(Q)$ at the same temperature.}     
\label{fig:7}
\end{figure}

At high temperatures, both $Q_i$ tend to be small. 
At low temperatures, the system spends most of its 
time close to one or the other of the structures FU and BU; 
either $\QFU$ or $\QBU$ is close to 1. Finally, at the collapse temperature, 
all three of these regions are populated, as can be seen 
from Figure~\ref{fig:7}a. In particular, this implies that  
folding and collapse occur at the same temperature. 

In Figure~\ref{fig:7}b, we show the free-energy profile $F(Q)$ at
the folding temperature, where $Q=\max(\QFU,\QBU)$ is taken as a measure of 
``nativeness''. The free energy has a relatively sharp minimum at 
$Q\approx0.9$, corresponding to $\delta=\min(\dFU,\dBU)\approx3$\AA. 
This is followed by a weak barrier around $Q=0.7$, 
corresponding to $\delta\approx6$\AA. Finally, there is a broad 
minimum at small $Q$, where $Q=0.2$ corresponds 
to $\delta\approx13$\AA. In \cite{Irback:00}, it was shown  
that the low-$Q$ minimum corresponds to expanded structures with a 
varying secondary-structure content.      

In particular, these results show that the three-helix sequence exhibits a  
first-order-like collapse transition that coincides with its folding 
transition. This is the behavior for $\kappa=0$ (see Equation \ref{kappa}). 
Next we discuss the character of the collapse transition for 
$\kappa\ne0$, starting with positive $\kappa$. 

\subsection{Dominant Hydrophobicity Forces}

A positive $\kappa$ means strong hydrophobicity forces and weak 
hydrogen bonds. For small positive $\kappa$, the collapse 
temperature remains approximately the same as for $\kappa=0$. 
However, the transition gets weaker with increasing $\kappa$. 
This is illustrated in Figures~\ref{fig:3}b and \ref{fig:4}, 
using data obtained for $\kappa=0.3$. 

In Figure~\ref{fig:3}b, we show the specific heat for $\kappa=0.3$. 
Compared to the $\kappa=0$ results (see Figure~\ref{fig:3}a), 
we see that the peak in the specific heat is lower, and that 
the chain-length dependence is weaker. There is no sign that 
the chain collapse is first-order-like for $\kappa=0.3$.

Consistent with the data for the specific heat, we see from  
Figure~\ref{fig:4} that the radius
of gyration changes more slowly with temperature 
for $\kappa=0.3$ than for $\kappa=0$.  

It is also interesting to look at the secondary-structure content. 
From Figure~\ref{fig:5}, it can be seen that the hydrogen-bond 
energy $\Ehb$ is considerably higher for $\kappa=0.3$ than for 
$\kappa=0$. In particular, the results show that the 
secondary-structure content at the collapse temperature 
is lower for $\kappa=0.3$.   

For $\kappa=0.3$, we furthermore find that the three-helix sequence 
does not show structural stability at temperatures immediately below 
the collapse transition (data not shown), so the folding temperature is 
different from and lower than the collapse 
temperature in this case. Between these two temperatures, 
the chain exists in a compact (molten globule) state
without specific structure.  
 
\subsection{Dominant Hydrogen Bonds}

We now turn to negative $\kappa$, meaning strong hydrogen bonds
and weak hydrophobicity forces. It is clear that the three-helix
sequence will form one long helix rather than a helical bundle if  
$\kappa$ is made too large negative. To get an idea of when  
this happens, we compare the energies of an optimized 
three-helix-bundle conformation and an optimized rodlike conformation,
for different $\kappa$. These conformations were generated as follows.

Starting at $\kappa=0$, we quenched a large number of 
low-temperature Monte Carlo conformations to zero
temperature, by using a conjugate-gradient method.  
The structure with the lowest energy found is the BU 
structure in Figure~\ref{fig:6}. This structure is
taken as our three-helix-bundle conformation at $\kappa=0$. 
Our rodlike $\kappa=0$ conformation was also obtained by 
a conjugate-gradient minimization, starting from a long ``ideal'' helix. 

\begin{figure}
\begin{center}
\epsfig{figure=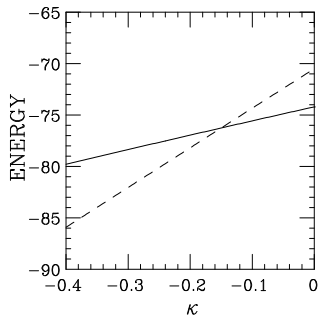,width=4.1cm,height=4.1cm}
\end{center}
\caption{Three-helix-bundle (full line) and one-helix (dashed line) 
energies against $\kappa$ (see the text).} 
\label{fig:8}
\end{figure}

We then performed energy minimizations at successively lower 
$\kappa$, each time taking the optimized conformations from the previous 
$\kappa$ as our two starting points. The two sets of energies  
obtained this way are shown as functions of $\kappa$ in Figure~\ref{fig:8}. 
We see that the curves cross at $\kappa\approx-0.15$. Although there
may well exist three-helix-bundle energies that are somewhat lower than those
in Figure~\ref{fig:8}, these results strongly suggest that the
ground state turns into one long helix already at a relatively small 
negative $\kappa$. 

\section{Summary}\label{disc}

The calculations discussed in this paper can be divided into two parts. 
First, we showed that the three-helix-bundle protein, for a suitable 
choice $(\tehb,\teaa)$ of the parameters $(\ehb,\eaa)$, indeed 
has the properties listed in the introduction. Let us stress that we
find these properties without resorting to the G\=o approximation.  
This is important as many current models rely on this approximation
\cite{Nymeyer:98,Shea:98,Shea:99,Clementi:00a,Clementi:00b,Shimada:01}, 
based on the view that the folding properties are strongly influenced
by the native topology, whereas energetic frustration plays a less 
important role. The results presented in this paper are consistent 
with this view, but it is clear that further studies are needed 
in order to properly understand the consequences and applicability of 
the G\=o approximation.            
 
In the second part, we presented results obtained for 
$(\ehb,\eaa)=(\tehb-\kappa,\teaa+\kappa)$ for different $\kappa$.
Not unexpectedly, it turns out that the folding behavior depends 
critically on the relative strength of the parameters $\ehb$ and $\eaa$. 
In particular, we saw that a first-order-like collapse 
to a three-helix-bundle state is observed only in a narrow 
window around $\kappa=0$; a proper balance between hydrogen bonds 
and hydrophobicity forces is required for the chain to show this
behavior. 

The fact that the dependence on these parameters is strong may seem 
unwanted, but is not physically unreasonable. In fact, the situation 
is somewhat reminiscent of what has been found for homopolymers with 
stiffness~\cite{Kolinski:86,Doniach:96,Bastolla:97,Doye:98}, 
with the hydrogen bonds playing the role of the stiffness term. 
Note also that the incorporation of full side chains will make the chains 
intrinsically stiffer, which might lead to a weaker dependence on  
the hydrogen-bond strength $\ehb$. 

\subsection*{Acknowledgements}
This work was in part supported by
the Swedish Foundation for Strategic Research.

\end{document}